# RingFormer: A Neural Vocoder with Ring Attention and Convolution-Augmented Transformer

Seongho Hong and Yong-Hoon Choi, *Member, IEEE*

*Abstract*—While transformers demonstrate outstanding performance across various audio tasks, their application to neural vocoders remains challenging. Neural vocoders require the generation of long audio signals at the sample level, which demands high temporal resolution. This results in significant computational costs for attention map generation and limits their ability to efficiently process both global and local information. Additionally, the sequential nature of sample generation in neural vocoders poses difficulties for real-time processing, making the direct adoption of transformers impractical. To address these challenges, we propose RingFormer, a neural vocoder that incorporates the ring attention mechanism into a lightweight transformer variant, the convolution-augmented transformer (Conformer). Ring attention effectively captures local details while integrating global information, making it well-suited for processing long sequences and enabling real-time audio generation. RingFormer is trained using adversarial training with two discriminators. It is integrated as a replacement for the decoder module in the VITS text-to-speech framework, enabling fair comparisons with state-of-the-art vocoders such as HiFi-GAN, iSTFT-Net, and BigVGAN under identical conditions using various objective and subjective metrics. Experimental results show that RingFormer achieves comparable or superior performance to existing models, particularly excelling in real-time audio generation. Our code and audio samples are available on GitHub.

## I. Introduction

AUDIO generation models have become core technologies in various application fields such as speech synthesis, music generation, and sound effect creation. Recent advancements have significantly enhanced generation quality and stability through generative adversarial network (GAN)-based models (e.g., Parallel WaveGAN [1], HiFi-GAN [2], BigVGAN [3], Avocodo [4]) and diffusion models (e.g., Grad-TTS [5], WaveGrad [6], Diff-TTS [7], E3 TTS [8]), both aiming to achieve high-quality speech synthesis.

Text-to-speech (TTS) models, which map text input to speech output, have seen major improvements in recent years by leveraging advancements in generative models. Among the components of a TTS system, vocoders play a pivotal role in determining the final audio quality. They are responsible for converting intermediate audio representations, such as mel-spectrograms, into waveform audio. A high-performing vocoder is essential for achieving natural and high-fidelity speech, as it directly impacts both the clarity and temporal consistency of the output audio. As vocoders are directly responsible for waveform generation, their design has a significant impact on the naturalness, clarity, and overall quality of synthesized speech.

GAN-based vocoders [1], [2], [3], [4] have emerged as a leading approach due to their ability to generate high-resolution speech in real-time. This capability makes them suitable for tasks such as TTS and speech restoration. However, GAN-based models face inherent challenges; while they produce sharp and detailed audio, they struggle with capturing long-term dependencies and complex patterns crucial for high-fidelity speech. Furthermore, training GAN models can be unstable, leading to mode collapse or inconsistencies in the generated audio. Despite these drawbacks, GAN-based vocoders remain a strong choice for real-time and high-resolution applications.

In contrast, diffusion models [5], [6], [7], [8] have gained attention for their ability to enhance the stability and quality of the audio generation process. By employing a step-by-step refinement process, diffusion models can produce consistent and natural-sounding speech, excelling in capturing complex and subtle audio details. This makes them particularly well-suited for high-quality, non-real-time synthesis. However, recent research has pointed out that these models may have limitations for time-sensitive applications due to slower generation speeds and higher computational demands.

In addition to GANs and diffusion models, flow-based models (e.g., WaveGlow [9], Flow-TTS [10], P-Flow [11], ReFlow-TTS [12]) and autoregressive models (e.g., Tacotron [13], NaturalSpeech [14]) have contributed to advancements in efficiency and quality. Autoregressive models excel at modeling the natural flow of speech but often sacrifice speed for quality. Flow-based models strike a balance between speed and fidelity but are less widely used than GANs and diffusion models in

Date of submission January 2, 2025. This work was supported by the Korea Agency for Infrastructure Technology Advancement (KAIA) grant funded by the Ministry of Land, Infrastructure and Transport under the Smart Building R&D Program (Grant No. RS-2025-02532980); by the Korea Institute for Advancement of Technology (KIAT) grant funded by the Ministry of Trade, Industry and Energy (MOTIE) under the HRD Program for Industrial Innovation (Grant No. RS-2024-00406796); and by the Research Grant of Kwangwoon University in 2023.

Corresponding author: Yong-Hoon Choi.
Seongho Hong and Yong-Hoon Choi are with the Fintech and AI Robotics (FAIR) Laboratory, the School of Robotics, Kwangwoon University, Nowon-gu, Seoul 01897, South Korea (e-mail: gyogook608@gmail.com; yhchoi@kw.ac.kr).

speech synthesis. Optimized architectures such as iSTFT-Net [15] have further improved real-time processing efficiency, and multimodal audio generation models leveraging inputs such as text, images, and video have opened new possibilities for innovative applications. Non-autoregressive approaches (e.g., FastSpeech [16], Parallel WaveGAN [1]) have also demonstrated significant strides in speed and quality, enabling real-time and interactive applications.

Despite these advancements, significant challenges persist. GAN-based vocoders are effective for generating high-resolution audio but still struggle with capturing long-term dependencies, which can lead to quality degradation. Diffusion models have improved stability but remain computationally expensive and unsuitable for real-time applications due to their sequential nature.

To address these challenges, we propose a novel GAN-based vocoder called RingFormer that incorporates convolution-augmented Transformers, known as Conformer [17], and an efficient ring attention [18] mechanism introduced in previous research. Importantly, RingFormer is designed as a drop-in replacement for the decoder module of the well-known text-to-speech model VITS [19]. By replacing the original decoder with RingFormer, we retain the overall probabilistic latent-variable modeling framework of VITS while enhancing the generation quality and efficiency at the waveform synthesis stage. While GANs offer the speed and high resolution necessary for real-time synthesis, RingFormer leverages the Conformer architecture to better capture both local details and global dependencies, addressing key weaknesses of traditional GAN-based models. Furthermore, ring attention enhances computational efficiency by focusing attention on localized regions while maintaining the ability to model long-range dependencies. This hybrid architecture, RingFormer, balances the trade-offs between speed and resolution, achieving the temporal resolution and efficiency needed for real-time speech synthesis while maintaining the high-quality audio output expected from modern TTS systems.

The remainder of this paper is organized as follows: Section II reviews related work. Section III describes the proposed model architecture, Section IV explains the loss functions, Section V presents experimental results and performance analysis, and Section VI concludes the paper.

## II. RELATED WORK

GANs have emerged as powerful models in the domain of audio synthesis, particularly for generating high-quality raw audio waveforms. WaveGAN [20], introduced by Donahue et al., was the first GAN-based approach designed to directly generate raw audio waveforms by adapting the DCGAN [21] architecture for one-dimensional audio data. Although WaveGAN demonstrated the feasibility of unsupervised learning for audio generation, it faced limitations in capturing fine-grained details. Building on this foundation, MelGAN [22] introduced a multiscale discriminator that leveraged average pooling to downsample audio at multiple scales. By incorporating win-dow-based discriminators to model audio features across different resolutions, MelGAN achieved efficient and high-quality audio synthesis with improved fidelity.

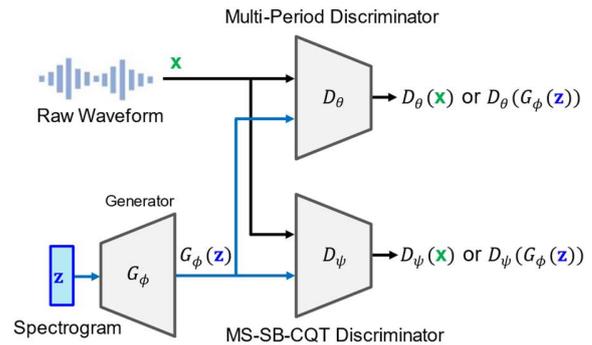

**FIGURE 1.** The overall structure of RingFormer. The waveform $G_\phi(z)$ generated by the generator and the real waveform x are input into two discriminators, each of which validates the quality of the generated speech in different ways.

HiFi-GAN [2], proposed by Kong et al., advanced the field by adopting a multi-period discriminator capable of capturing periodic structures in time-domain audio. The model combined short-time Fourier transform (STFT) loss and mel-spectrogram loss, enabling it to generate high-resolution, natural-sounding audio suitable for speech synthesis and restoration tasks. GAN-TTS [23] further refined the use of GANs in audio synthesis by utilizing a conditional feed-forward generator alongside an ensemble of discriminators that operated on random windows of varying sizes. This approach enabled GAN-TTS to achieve high-quality audio synthesis while maintaining both local coherence and global consistency.

Parallel WaveGAN [1], introduced by Yamamoto et al., incorporated a combination of multi-resolution STFT loss and adversarial loss in the waveform domain. This innovation allowed for parallel waveform generation, eliminating the need for complex probability density distillation techniques and significantly enhancing both generation speed and quality. Similarly, iSTFT-Net [15] simplified the output layers of traditional CNN-based vocoders by replacing them with inverse STFT layers. This design reduced model complexity and computational costs while maintaining audio quality.

BigVGAN [3], developed by Lee et al., pushed the boundaries of GAN-based audio synthesis by incorporating periodic activation functions to stabilize training and anti-aliasing techniques to reduce artifacts. These features enhanced fidelity and robustness in the generated audio, making BigVGAN a notable advancement in high-resolution audio synthesis.

While these GAN-based models have driven significant advancements in audio generation, they often struggle to capture long-term dependencies due to their reliance on iterative upsampling processes to expand receptive fields. This limitation can result in inconsistencies when modeling extended temporal relationships in audio data. To address these challenges, we propose a novel generator architecture, RingFormer, which integrates self-attention mechanisms with convolutional layers. This hybrid approach enables the model to effectively capture long-term dependencies while maintaining computational efficiency. Additionally, the incorporation of ring attention reduces computational overhead by focusing on fixed local regions, preserving both local and global relationships. Enhanced loss functions are also introduced to enable more accurate and efficient audio synthesis.

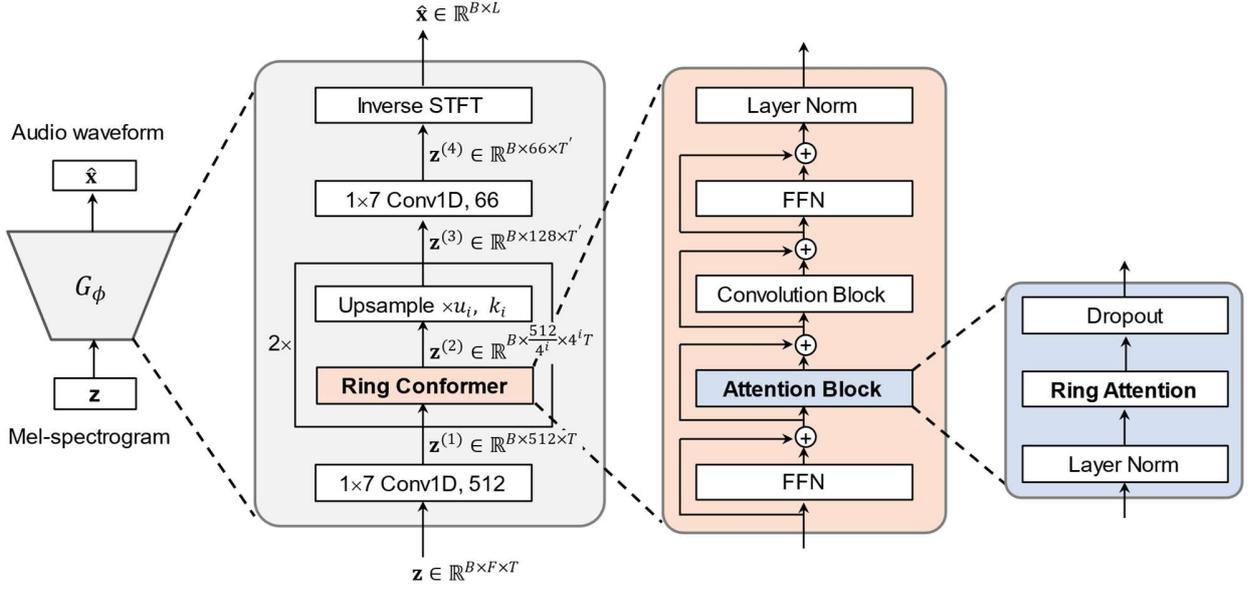

**FIGURE 2.** The overall structure of the RingFormer generator. The input mel-spectrogram to the RingFormer generator has the shape $B \times F \times T$, where $B$ is the batch size, $F$ is the number of bins, and $T$ and $T'$ are the number of time frames before and after upsampling, respectively. The kernel size $k_i$ of the transpose convolution for upsampling is 8, and the upsampling rate $u_i$ is 4.

### III. METHOD

The overall architecture of the proposed model consists of one generator and two discriminators, as shown in Figure 1. The generator maps the spectrogram $z$ to an audio waveform $G_\phi(z)$, while the two discriminators $D_\theta$ and $D_\psi$ compare the real audio waveform $x$ and the generated waveform $G_\phi(z)$ in different ways.

#### A. Generator

Recognizing that capturing long-term dependencies is crucial for modeling realistic speech audio, we propose a new generator architecture designed to learn these dependencies more effectively. The proposed architecture, as shown in Figure 2, incorporates two stages of Conformer blocks with ring attention and ×4 upsampling between the input and output convolutions. This approach contrasts with the upsampling process in HiFi-GAN, which uses the multi-receptive field fusion (MRF) technique with [×8, ×8, ×2, ×2] transpose convolutions to perform upsampling and reconstruct raw audio. In comparison, our proposed structure simplifies the upsampling process by using two stages of Conformer blocks with ring attention and ×4 upsampling, providing a more efficient and streamlined approach to generating high-quality audio. The remaining Conformer blocks, excluding the ring attention, are identical to those in [18]. This modification improves the ability to capture long-term dependencies in the generated audio, enhancing the model's overall performance and synthesis quality.

In the upsampling block, the snake activation function [24] helps the model learn the periodic structure of speech signals more accurately. Although the final output of the generator is the magnitude and phase of the spectrogram rather than the waveform, these components also exhibit periodic characteristics, making them suitable for modeling periodic structures. Unlike BigVGAN [3], no anti-aliasing filter is used for upsampling, as smaller upsampling ratios allow for more stable high-frequency processing. After upsampling, the inverse STFT reconstructs the signal in the frequency domain, separating amplitude and phase for better control. This structure maintains memory efficiency for long sequences while improving the learning of long-term dependencies in speech signals. Through these improvements, RingFormer achieves more precise speech synthesis without sacrificing speed.

#### B. Ring Attention

Capturing long-term dependencies is crucial for modeling realistic speech audio. For instance, the duration of a phoneme can exceed 100ms, resulting in a high correlation between more than 2,200 adjacent samples in the raw waveform. Ring attention [18] is a mechanism designed to efficiently process long sequences by leveraging blockwise parallel computation. In RingFormer, ring attention is tailored for vocoders to effectively handle long sequences of speech signals.

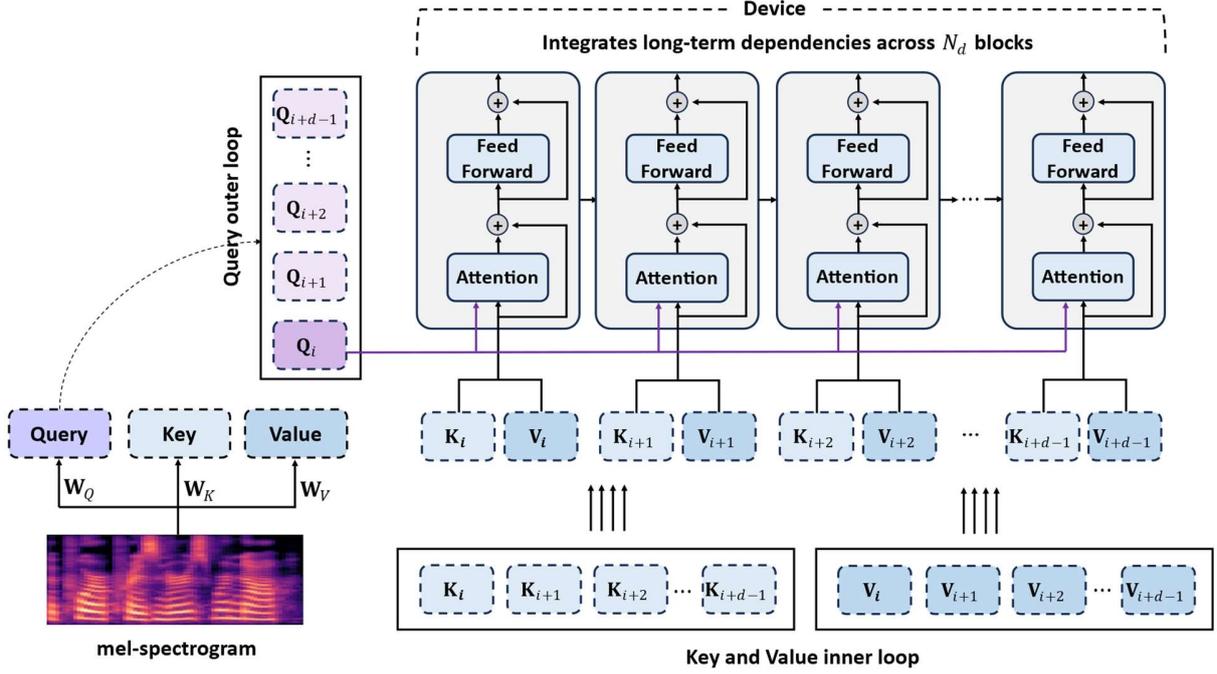

**FIGURE 3.** The operation process of Ring Attention.

First, the mel-spectrogram upsampled from the MRF is divided into $N_d$ fixed-size blocks, and each block is assigned to an individual device. Here, device refers to a physical GPU used for blockwise parallel computation in the Ring Attention mechanism. Each device generates query, key, and value based on the divided mel-spectrogram, which are obtained through affine transformations using learnable weight matrices $\mathbf{W}_Q$, $\mathbf{W}_K$, and $\mathbf{W}_V$.

Subsequently, a key-value exchange mechanism based on a ring topology allows each device to receive key and value data from its adjacent device. This data exchange enables information to flow between blocks, thereby effectively integrating global dependencies and context across the entire sequence. This structure is well-suited for modeling both the temporal dependencies of speech signals and the harmonic structure within frequency bands, allowing it to capture the periodic characteristics of speech in detail.

In particular, ring attention effectively resolves the memory bottleneck issue encountered when processing long sequences in vocoders. Since key-value exchanges and attention computations are designed to be performed in parallel across devices, computational efficiency is maximized, significantly reducing memory and computational costs during the training process for long sequences of data. Blockwise attention computations within the device are carried out as follows:

$$\text{Attention}_i(Q_i, K, V) = \text{softmax}\left(\frac{Q_i K^T}{\sqrt{d_k}}\right) V, \quad (1)$$

where $i$ denotes the device index, and $d_k$ represents the dimension of the key vector. The query $Q_i$ performs a scaled dot-product computation with the keys $K = \{K_i, ..., K_{i+d-1}\}$ in the same device, which is then multiplied with $V = \{V_i, ..., V_{i+d-1}\}$ to calculate the attention values.

This method overcomes the memory constraints of traditional Transformer [25] models, allowing the context size to scale linearly with the number of devices. As a result, ring attention maintains computational efficiency while achieving high performance in both training and inference for extremely large context sizes.

*C. Discriminators*

We use two discriminators for generator training: the multi-period discriminator (MPD) and the multi-scale sub-band constant-Q transform (MS-SB-CQT) discriminator.

Since speech audio consists of sinusoidal signals with various periods, it is necessary to identify the diverse periodic patterns inherent in the audio data. To this end, HiFi-GAN [2] proposed the MPD, and in this paper, we use the same MPD without modification.

Additionally, the MS-SB-CQT discriminator [26] improves upon the multi-scale discriminator (MSD) of Mel-GAN [22] by using constant-Q transform (CQT) to process more precise frequency band information. This approach enhances both frequency and time resolution, capturing more detailed characteristics of the speech signal and enabling more natural speech synthesis results. While the original MSD focused on capturing information across multiple frequency ranges, CQT allows for more detailed frequency band analysis, providing finer frequency interpretation. In this paper, the MS-SB-CQT discriminator is used without modification.

The use of MPD and MS-SB-CQT discriminators in our approach is largely consistent with common practices in recent GAN-based vocoder models. These discriminators are employed without architectural modifications, following conventions established in HiFi-GAN and subsequent works. Therefore, our use of discriminators aligns with standard techniques in the literature rather than introducing novel discriminator designs.

## IV. TRAINING OBJECTIVE DETAILS

We use various loss functions to optimize RingFormer. To evaluate speech quality, it is integrated into the widely used TTS model VITS [19] and trained by connecting it to the model. As a result, the encoder parameters of VITS are also updated.

### A. Adversarial Loss

The RingFormer is trained using two discriminators. The first is the MPD, originally proposed in HiFi-GAN [2], and the second is the MS-SB-CQT discriminator [26]. The MPD is designed as a combination of sub-discriminators based on Markovian windows, with each sub-discriminator specializing in detecting different periodic patterns in the input waveform. This structure allows for a systematic evaluation of speech data with diverse periodic characteristics. However, a limitation of the MPD is that its sub-discriminators evaluate only isolated samples, potentially overlooking broader contextual information. To address this limitation, the MS-SB-CQT discriminator, as proposed in [26], is incorporated to enhance performance. The adversarial loss is defined as follows:

$$\mathcal{L}_{adv} = \mathcal{L}_G + \mathcal{L}_D, \tag{2}$$

$$\mathcal{L}_G = \alpha \mathbb{E}_z \left[ \left(1 - D_\theta \left(G_\phi(z)\right)\right)^2 \right] \\ + (1-\alpha) \mathbb{E}_z \left[ \left(1 - D_\psi \left(G_\phi(z)\right)\right)^2 \right], \tag{3}$$

$$\mathcal{L}_D = \alpha \mathbb{E}_{x,z} \left[ \left(1 - D_\theta(x)\right)^2 + \left(D_\theta \left(G_\phi(z)\right)\right)^2 \right] \\ + (1-\alpha) \mathbb{E}_{x,z} \left[ \left(1 - D_\psi(x)\right)^2 \\ + \left(D_\psi \left(G_\phi(z)\right)\right)^2 \right]. \tag{4}$$

The contribution of each discriminator to the training loss is controlled by a weighting factor, $\alpha$, which is set to 0.5 to balance their roles during adversarial training.

### B. Spectral Decomposition Loss

In our work, we explicitly model both magnitude and phase losses, following the findings in [27]. This decomposition enables independent optimization of spectral energy (magnitude) and phase structure, leading to more accurate waveform reconstruction. Specifically, the magnitude loss promotes faithful spectral amplitude reproduction, while the phase loss enhances temporal consistency by aligning fine-grained phase information. This joint optimization achieves a balanced trade-off between time and frequency domains, thereby improving perceptual quality and generalization across diverse audio conditions. The spectral decomposition loss is defined as follows:

$$\mathcal{L}_{sd} = \mathcal{L}_{mag} + \mathcal{L}_{phase}, \tag{5}$$

$$\mathcal{L}_{mag} = \mathbb{E}_{x,z} \left[ \left\| |F(x)| - |F(G_\phi(z))| \right\|_1 \right], \tag{6}$$

$$\mathcal{L}_{phase} = \mathbb{E}_{x,z} \left[ 1 - \mathrm{Re} \left( \frac{F(x)}{|F(x)|} \cdot \frac{F(G_\phi(z))^*}{|F(G_\phi(z))|} \right) \right]. \tag{7}$$

Here, $F(\cdot)$ denotes the short-time Fourier transform (STFT), and $(\cdot)^*$ denotes the complex conjugate. The phase loss term computes the cosine distance between normalized complex spectra, effectively addressing the phase wrapping issue and ensuring robust alignment of phase information in the frequency domain.

### C. Feature Matching Loss

The feature matching loss $\mathcal{L}_{fm}$ [22] minimizes the $\ell_1$ distance between the intermediate features extracted from the discriminator layers:

$$\mathcal{L}_{fm} = \mathbb{E}_{x,z} \left[ \sum_{i=1}^{T} \frac{1}{N_i} \left\| D_k^i(x) - D_k^i \left(G_\phi(z)\right) \right\|_1 \right], \tag{8}$$

where $T$ is the number of layers in the sub-discriminator $D_k$, and $N_i$ is the number of features in the $i$-th layer. The feature matching loss encourages the generator to produce outputs whose intermediate features are similar to those of the real data, improving the generator's ability to match the discriminator's learned feature representations.

### D. Final Loss

The proposed RingFormer is implemented to replace the decoder of the widely used end-to-end TTS model, VITS. While the training environment is integrated with VITS, RingFormer is not dependent on it. Unlike models such as FastSpeech [16], VITS eliminates the need for a separate duration predictor or aligner (e.g., attention alignment in Tacotron [13]). Additionally, VITS combines a GAN with a variational autoencoder (VAE) [28] to generate high-resolution and natural-sounding speech.

In this paper, the proposed RingFormer is optimized using two additional loss functions adopted from VITS. The first is $\mathcal{L}_{dur}$, which facilitates learning text-to-speech alignment, and the second is $\mathcal{L}_{KL}$, which plays a critical role in modeling the relationship between text and speech in the latent space. $\mathcal{L}_{KL}$ regulates the distribution of latent variables, enabling natural speech synthesis and supporting the alignment-free structure. These two loss functions are applied without modification during the training of the proposed model. The total loss function is defined as follows:

$$\mathcal{L}_{total} = \mathcal{L}_{adv} + \lambda_{sd}\mathcal{L}_{sd} + \lambda_{fm}\mathcal{L}_{fm} + \lambda_{recon}\mathcal{L}_{recon} + \lambda_{KL}\mathcal{L}_{KL} + \lambda_{dur}\mathcal{L}_{dur}. \quad (9)$$

The hyperparameters $\lambda_{sd}, \lambda_{fm}, \lambda_{recon}, \lambda_{KL}, \lambda_{dur}$ were set to 0.7, 1, 45, 1, and 1, respectively, in this study. These values were determined empirically to balance the contributions of each loss component, considering their relative magnitudes and impact on training stability and performance. Unlike assigning uniform weights, this configuration allows more emphasis to be placed on the reconstruction loss, which was observed to play a dominant role in optimizing audio quality, while maintaining the effects of the other loss terms.

## V. EXPERIMENTS

To validate the performance of RingFormer, it is applied to the decoder of the widely used TTS model, VITS [19], and the quality of the synthesized speech is evaluated. For comparison, the baseline vocoders used are HiFi-GAN [2], iSTFT-Net [15], and BigVGAN [3], which are state-of-the-art models known for achieving top performance in the field. These models are also applied to VITS with the same architecture and hyperparameters to ensure a fair comparison under equal conditions.

In addition, considering that RingFormer adopts a spectral modeling approach by employing iSTFT and operating in the frequency domain, we further compare our model with recent spectral modeling-based neural vocoders, including Vocos [29] and APNet [30]. Both models have demonstrated strong performance with efficient spectral representations. These vocoders are also integrated into the same VITS architecture for consistency, and their performance is reported alongside other baselines. Our code and audio samples are available on GitHub [31].

### A. Dataset
We trained and evaluated the model using the widely adopted LJSpeech dataset [32], which contains 13,100 high-quality English speech samples totaling approximately 24 hours, recorded at a sampling rate of 22,050 Hz. From this dataset, 12,500 samples were used for training, 500 for validation, and 30 samples were randomly selected from the remaining data for final testing. This configuration ensured consistent and reliable model evaluation throughout the study.

### B. Experimental Setup
The proposed generator in this study employs two stages of upsampling. Initially, the number of channels is set to 512, and at each stage, the number of channels is halved according to $2^i$, where $i$ denotes the upsampling step. The Conformer block is configured by adjusting the input dimensions at each upsampling stage. It utilizes 8 attention heads, a feed-forward network dimension that is half the input dimension, 2 Conformer layers, a kernel size of 31, and a dropout rate of 0.1. Key hyperparameters are summarized in Table 1.

**TABLE 1. Hyperparameters of RingFormer generator and discriminators**

| Layer | Hyperparameters | Values |
|---|---|---|
| Generator | Upsample rates ($u_i$) | [4, 4] |
| | Upsample kernel size ($k_i$) | [8, 8] |
| | Number of input channels ($h$) | 512 |
| | Number of output channels | 66 |
| | iSTFT filter size | 64 |
| | iSTFT hop size | 16 |
| | Dropout rate | 0.1 |
| | Ring sequence length | 512 |
| | Attention head dimension | 64 |
| | Number of Attention heads | 8 |
| | Batch size | 64 |
| MS-SB-CQT Discriminator | Hop lengths | [512, 256, 256] |
| | Number of octaves | [9, 9, 9] |
| | Bins per octaves | [24, 36, 48] |
| Multi-Period Discriminator | Periods | [2, 3, 5, 7, 9] |
| | Kernel size | 5 |
| | Stride | 3 |

Model training was conducted on an Ubuntu 20.04 LTS system using PyTorch 2.4.1 and CUDA 12.2. Software dependencies were managed through Docker. The training was performed on a server equipped with four NVIDIA A100 GPUs (80 GB HBM2 memory each, connected via NVLink), of which two GPUs were allocated for training the proposed model.

### C. Evaluation Metrics
In this study, the model's performance is evaluated from multiple perspectives using mel-cepstral distortion (MCD) [33], word error rate (WER), short-time objective intelligibility (STOI) [34], NISQA [35], mean opinion score (MOS), and comparison MOS (CMOS).

MCD measures the difference in mel-frequency cepstral coefficients between the synthesized and reference speech. A lower MCD value indicates higher similarity between the two voices. WER evaluates the accuracy of speech recognition by measuring the alignment of the recognized text with the original transcript. A lower WER indicates fewer recognition errors. STOI quantifies the intelligibility of the synthesized speech in relation to the reference, with values ranging from 0 to 1. A value closer to 1 indicates higher intelligibility. NISQA is a deep learning-based metric for assessing the quality and naturalness of speech by mimicking human auditory perception and quantifying the subjective quality of speech.

MOS is a subjective evaluation metric, where listeners rate the speech quality on a scale from 1 to 5. However, in this study,

**TABLE 2.** The performance of RingFormer and the baseline models evaluated on LJSpeech. MOS is objectively evaluated on a scale from 1 (very unpleasant) to 5 (very satisfactory), while CMOS is subjectively evaluated on a scale from -3 (very bad) to 3 (very good). Values in ( ) are 95% confidence intervals. Bold: the best, Blue: the 2nd best.

| Model | MCD ↓ | WER ↓ | STOI ↑ | NISQA ↑ | MOS ↑ | CMOS ↑ |
|---|---|---|---|---|---|---|
| Ground Truth | - | 0.048 (±0.01) | - | 4.644 (±0.04) | 4.35 (±0.01) | - |
| HiFi-GAN | 0.328 (±0.04) | 0.076 (±0.01) | 0.972 (±0.05) | 4.344 (±0.03) | 3.93 (±0.02) | -0.182 (±0.13) |
| iSTFT-Net | 0.322 (±0.04) | 0.073 (±0.01) | 0.969 (±0.05) | 4.312 (±0.03) | 3.89 (±0.02) | -0.180 (±0.10) |
| BigVGAN | **0.311 (±0.05)** | **0.066 (±0.01)** | 0.983 (±0.02) | **4.421 (±0.04)** | 4.03 (±0.02) | -0.169 (±0.10) |
| APNet | 0.326 (±0.09) | 0.077 (±0.01) | 0.980 (±0.03) | 4.369 (±0.06) | 3.99 (±0.03) | -0.179 (±0.16) |
| Vocos | 0.319 (±0.03) | 0.069 (±0.01) | 0.974 (±0.03) | 4.397 (±0.03) | 4.00 (±0.04) | -0.171 (±0.12) |
| RingFormer | 0.314 (±0.03) | 0.068 (±0.01) | **0.985 (±0.02)** | 4.410 (±0.02) | **4.09 (±0.02)** | **-0.162 (±0.15)** |

**TABLE 3.** Comparison of model size and inference speed. Inference speed is the relative speed compared to real-time using a GPU.

| Model | # Param (M) | Speed on GPU | Speed on CPU | Decoding Cost (GMAC/s) | Peak Memory (MB) |
|---|---|---|---|---|---|
| HiFi-GAN | 14.33 | ×182.46 | ×4.02 | 319.63 | 339.4 |
| iSTFT-Net | 13.66 | ×194.39 | ×7.84 | 225.39 | 334.3 |
| BigVGAN | 114.80 | ×93.65 | ×1.18 | 871.18 | 1114.8 |
| APNet | 72.19 | ×201.03 | ×9.89 | 64.95 | 918.2 |
| Vocos | 13.53 | ×233.91 | ×18.32 | 12.19 | 201.9 |
| RingFormer (w/ vanilla attention) | 30.30 | ×99.02 | ×4.35 | 95.20 | 2299.3 |
| RingFormer | 30.10 | ×186.87 | ×9.45 | 99.28 | 374.5 |

we utilize the MOS prediction system from UTMOS [36] for objective evaluation, which produces scores highly correlated with human ratings. Finally, CMOS is used to compare the relative quality between two speech samples, where one is always the ground truth reference. A total of 60 volunteer listeners participated in the evaluation of 210 audio samples, each comparing a synthesized sample against its corresponding ground truth on a 7-point scale ranging from –3 (much worse) to +3 (much better). This large-scale subjective evaluation provides additional insight into the perceptual effectiveness of the proposed vocoder. These diverse metrics enable a comprehensive evaluation of the model's speech quality, intelligibility, and naturalness.

*D. Results*

We report the performance evaluation results of RingFormer and baseline vocoder models on LJSpeech using the previously described objective and subjective metrics, as summarized in Table 2.

The proposed model demonstrates overall stable performance, achieving particularly strong results in MOS, which evaluates the naturalness and quality of speech, surpassing other models. It also maintains consistent quality in the NISQA metric, indicating reliable generation without perceptual artifacts. While slightly behind BigVGAN in MCD and STOI, RingFormer achieves comparable results, confirming its competitiveness in terms of speech similarity and intelligibility. In WER, which measures pronunciation accuracy, the model performs on par with other baselines. Furthermore, it shows marginally better CMOS scores than BigVGAN and outperforms HiFi-GAN.

Compared to lightweight spectral modeling-based vocoders such as Vocos and APNet, RingFormer consistently delivers superior quality across multiple evaluation metrics, demonstrating its effectiveness and practicality for high-quality speech synthesis applications.

Table 3 presents a comparison of model size, inference speed, decoding cost, and peak memory usage for the proposed RingFormer and baseline models. The proposed model contains 30.10M parameters and achieves a GPU inference speed of approximately ×186.87, with a decoding cost of 99.28 GMAC/s and a peak memory usage of 374.5 MB. Compared to BigVGAN, RingFormer reduces decoding cost by roughly 8.8 times and peak memory usage by about 3 times, demonstrating significantly improved efficiency. Additionally, it outperforms HiFi-GAN and iSTFT-Net in decoding efficiency, achieving reductions of 3.2× and 2.3× respectively, while maintaining a comparable level of memory usage, which highlights its strong practicality in resource-constrained environments.

**TABLE 4.** Ablation results evaluated on LJSpeech. 'w/o all' denotes the model trained without the magnitude loss ($\mathcal{L}_{mag}$), phase loss ($\mathcal{L}_{phase}$), and the MS-SB-CQT discriminator.

| Model | MCD ↓ | WER ↓ | STOI ↑ | NISQA ↑ | MOS ↑ |
|---|---|---|---|---|---|
| RingFormer | **0.314 (±0.03)** | 0.068 (±0.01) | **0.985 (±0.02)** | 4.410 (±0.02) | 4.09 (±0.01) |
| w/o Magnitude and Phase loss | 0.318 (±0.02) | 0.069 (±0.01) | 0.980 (±0.02) | 4.399 (±0.03) | 4.03 (±0.02) |
| w/o MS-SB-CQT discriminator | 0.320 (±0.02) | 0.071 (±0.01) | 0.978 (±0.03) | 4.405 (±0.03) | 4.05 (±0.02) |
| w/o all | 0.325 (±0.03) | 0.073 (±0.01) | 0.974 (±0.04) | 4.396 (±0.04) | 3.98 (±0.02) |
| w vanilla self-attention | 0.327 (±0.03) | 0.074 (±0.02) | 0.978 (±0.03) | 4.391 (±0.04) | 3.99 (±0.02) |
| w Ring sequence size 1024 | 0.316 (±0.02) | 0.073 (±0.02) | 0.983 (±0.02) | 4.408 (±0.04) | 4.08 (±0.01) |
| w Ring sequence size 256 | 0.315 (±0.03) | 0.067 (±0.01) | 0.985 (±0.04) | 4.417 (±0.02) | 4.14 (±0.01) |
| w Ring sequence size 128 | 0.324 (±0.02) | 0.071 (±0.02) | 0.976 (±0.01) | 4.401 (±0.03) | 4.06 (±0.02) |

**TABLE 5.** Pearson correlation between the F0 contour generated by the model and the ground truth.

| Model | Pearson Correlation |
|---|---|
| HiFi-GAN | 0.9142 |
| iSTFT-Net | 0.9129 |
| BigVGAN | 0.9176 |
| RingFormer | **0.9185** |

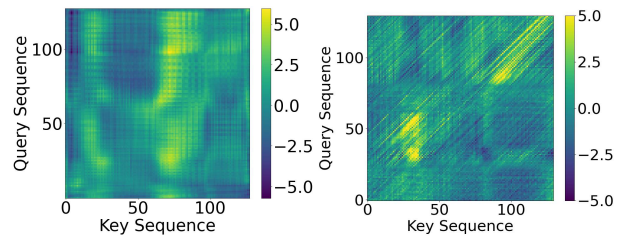

**FIGURE 4.** Attention maps extracted from the RingFormer model during the synthesis of the sentence "the window was approximately one hundred twenty feet away." The maps show the first segment of the input sentence, with the segment size set to 32 tokens as in training. (Left) Attention map using Ring Attention. (Right) Attention map using vanilla self-attention.

Notably, the ablation model without ring attention ("RingFormer w/ vanilla attention"), which corresponds to a standard Conformer architecture, results in a dramatic increase in peak memory usage to 2299.3 MB—nearly 6 times higher—underscoring the crucial role of ring attention in enabling memory-efficient inference.

These results collectively indicate that RingFormer is a competitive model capable of generating high-quality speech while meeting real-time processing requirements, making it well-suited for practical deployment in various applications.

Table 4 presents the ablation results that evaluate the contribution of each component in RingFormer to the quality of synthesized speech. The complete RingFormer model achieves the highest MOS of 4.09. Removing both the magnitude loss ($\mathcal{L}_{mag}$) and phase loss ($\mathcal{L}_{phase}$) leads to a noticeable performance degradation, reducing the MOS to 4.03. This suggests that both losses play a crucial role in capturing fine-grained periodicity and amplitude characteristics of speech. Similarly, eliminating the MS-SB-CQT discriminator also lowers performance (MOS 4.05), confirming the benefit of adversarial learning across time-frequency sequences. When all three components are removed, the MOS drops to 3.98, the lowest among the variants.

To further analyze the effectiveness of Ring Attention beyond speed, we include additional ablations on its architectural variants. Replacing Ring Attention with standard vanilla self-attention leads to degraded performance (MOS 3.99), demonstrating that Ring Attention contributes not only to efficiency but also to generation quality. Moreover, we vary the Ring sequence size to observe its effect on receptive field and modeling long-term dependencies. The best performance (MOS 4.14) is achieved when the Ring size is 256 (approximately 0.3 ms temporal window). Increasing the size to 1024 or decreasing it to 128 leads to reduced performance, indicating that appropriate localization of attention is essential. These findings underscore the importance of appropriately localized attention. Excessively global attention may obscure critical temporal structures, ultimately degrading model performance. This further supports the effectiveness of Ring Attention in balancing computational efficiency with high-quality speech generation.

Table 5 evaluates the ability of the RingFormer model to capture long-term dependencies by analyzing the autocorrelation of the F0 contour. The results show that the RingFormer model achieves the highest Pearson correlation coefficient with the ground truth, highlighting its strong performance in learning long-term dependencies.

Figure 4 presents a comparison between the attention score maps generated by the proposed Ring Attention mechanism (left) and the conventional vanilla self-attention (right) during speech synthesis. The Ring Attention map exhibits smoother and more spatially coherent patterns, characterized by gradual transitions across neighboring tokens. This indicates enhanced modeling of local context and improved temporal consistency, which are critical for generating natural and stable speech waveforms. In contrast, the attention map from vanilla self-attention displays irregular and high-frequency fluctuations, with abrupt changes in attention weights. Such patterns suggest unstable alignment behavior, which may negatively affect the continuity and naturalness of the synthesized audio. These visual differences highlight the effectiveness of Ring Attention in maintaining consistent and context-aware attention over long sequences, a property particularly beneficial in neural vocoding tasks.

## VI. CONCLUSION

In this paper, we propose RingFormer, a vocoder that efficiently processes long sequences with long-term dependencies through a Conformer block with Ring Attention, while maintaining a reasonable memory usage to synthesize high-quality speech. This structure captures both local and global dependencies in speech signals, enabling the generation of more natural-sounding speech. Additionally, to improve generation speed, the output layer incorporates an inverse STFT structure, and by adding phase and magnitude losses to the loss function, it finely learns temporal patterns and amplitude information, thereby enhancing the quality of the synthesized speech. For adversarial training, we introduce the recently released MS-SB-CQT discriminator, which improves the precision of speech synthesis by more accurately evaluating continuous sequences. Using various objective metrics (MCD, WER, STOI, and NISQA) and subjective evaluations (MOS and CMOS), we verify that Ring-Former performs on par with or better than existing models, achieving both naturalness and clarity in synthesized speech. This study presents a model that balances fast speech synthesis speed and high quality, contributing to the advancement of speech synthesis technology. Future research will aim to expand the applicability of RingFormer by optimizing it for multilingual datasets and various application environments.

ACKNOWLEDGMENT

The experiments in this paper were conducted using GPU resources provided by the "Convergence Open Shared System" Project, supported by the Ministry of Education and the National Research Foundation of Korea.